# Special polynomials and soliton dynamics

Yair Zarmi

Jacob Blaustein Institutes for Desert Research
Ben-Gurion University of the Negev
Midreshet Ben-Gurion, 84990, Israel

Abstract
Special polynomials play a role in several aspects of soliton dynamics.  These are differential polynomials in $u$, the solution of a nonlinear evolution equation, which vanish identically when $u$ represents a single soliton.  Local special polynomials contain only powers of $u$ and its spatial derivatives.  Non-local special polynomials contain, in addition, non-local entities (e.g., $\partial_x^{-1} u$).  When $u$ is a multiple-solitons solution, local special polynomials are localized in the vicinity of the soliton-collision region and fall off exponentially in all directions away from this region.  Non-local ones are localized along soliton trajectories.

Examples are presented of how, with the aid of local special polynomials, one can modify equations that have only a single-soliton solution into ones, which have that solution as well as, at least, a two-solitons solutions.

Given an integrable equation, with the aid of local special polynomials, it is possible to find all evolution equations in higher scaling weights, which share the same single-soliton solution and are either integrable, or, at least, have a two-solitons solution.  This is demonstrated for one or two consecutive scaling weights for a number of known equations.

In the study of perturbed integrable equations, local special polynomials are responsible for inelastic soliton interactions generated by the perturbation in the multiple-soliton case, and for the (possible) loss of asymptotic integrability.  Non-local special polynomials describe higher-order corrections to the solution, which are of an inelastic nature.

PACS: 02.30.IK, 02.30.Mv, 05.45.-a

## 1. Special polynomials – Definition and properties

Special polynomials are differential polynomials in $u$, the solution of a nonlinear evolution equation, which vanish when $u$ is a single-soliton solution. They are local if they contain only powers of $u$ and $\partial_x^k u$. They are non-local, if they also contain non-local entities, e.g., $q^{(1,1)} \equiv \partial_x^{-1} u$. Like evolution equations, special polynomials may be classified according to scaling weight, $W$. Consider, for example, the KdV equation:

$$u_t = S_2^{KdV}[u] = 6uu_1 + u_3 \qquad \left(u_k \equiv \partial_x^k u\right), \tag{1}$$

Assigning weights (2, 3, 1) to $u$, $\partial_t$ and $\partial_x$, respectively, Eq. (1) has $W = 5$ [1-8]. (The lowest symmetry, $S_1 = \partial_x u$, has $W = 3$.)

The single-soliton solution is given by:

$$u_{Single}[k] = \left(2k^2\right)/\left\{\cosh[k(x+vt)]\right\}^2 . \tag{2}$$

Special polynomials exist in the KdV case for $W \geq 3$. There are a finite number of linearly independent special polynomials in every $W$. They will be denoted by $R^{(W,l)}$. Possible choices for the linearly independent $R^{(W,l)}$ in $3 \leq W \leq 5$ (all non-local!) are:

$\underline{W = 3}$
$$R^{(3,1)} = u_1 + q^{(1,1)}u \quad, \quad R^{(3,2)} = 3u_1 + 3q^{(3,1)} - \left(q^{(1,1)}\right)^3, \tag{3}$$

$\underline{W = 4}$
$$R^{(4,1)} = \partial_x R^{(3,1)} \quad, \quad R^{(4,2)} = q^{(1,1)} R^{(3,1)} \quad, R^{(4,3)} = (1/3)\partial_x R^{(3,2)}$$
$$R^{(4,4)} = 2q^{(4,1)} + u^2 \tag{4}$$

$\underline{W = 5}$
$$R^{(5,1)} = \partial_x R^{(4,1)} \quad, \quad R^{(5,2)} = q^{(1,1)} R^{(4,1)}$$
$$R^{(5,3)} = \partial_x R^{(4,2)} \quad, \quad R^{(5,4)} = q^{(1,1)} R^{(4,2)}$$
$$R^{(5,5)} = \partial_x R^{(4,3)} \quad, \quad R^{(5,6)} = q^{(1,1)} R^{(4,3)} \quad, \quad R^{(5,7)} = q^{(1,1)} R^{(4,4)} . \tag{5}$$
$$R^{(5,8)} = q^{(5,2)} - q^{(5,3)} \quad, \quad R^{(5,9)} = 3q^{(5,1)} - 6q^{(5,3)} + q^{(1,1)} u_2 - u_3$$
$$R^{(5,10)} = 20 q^{(5,1)} - 55 q^{(5,3)} + \left(q^{(1,1)}\right)^5 - 5u_3$$

$q^{(W,l)}$ are bounded non-local entities of scaling weight $W$. Some are listed Appendix I. The construction of special polynomials is demonstrated in Appendix II.

A local special polynomial appears for the first time in $W = 6$, and is given by

$$R^{(6,1)} = u^3 - \left(u_1\right)^3 + uu_2 . \tag{6}$$

When $u$ is a multiple-solitons solution, local $R^{(W,l)}$ are appreciable only in the soliton-collision region, and fall off exponentially in all directions in the $x - t$ plane [35, 36]. (See, e.g., Fig.1.)

In contrast, non-local $R^{(W,l)}$ are localized along soliton trajectories. They are of a purely inelastic nature: Away from the soliton collision region, a non-local special polynomial tends asymptotically into a linear combination of the individual solitons, the amplitude of each affected only by the wave numbers of the *other* solitons. For example, when $u$ is a 3-solitons solution, the asymptotic form of $R^{(3,1)}$ of Eq. (3) is:

$$R^{(3,1)}[u] \underset{t \to \pm\infty}{\to} \pm 2\{(k_2 + k_3)u_{Single}[k_1] + (k_3 - k_1)u_{Single}[k_2] - (k_1 + k_2)u_{Single}[k_3]\} \ . \tag{7}$$

In every scaling weight $W$, some $R^{(W,l)}$, but not all, can be constructed by the application of scaling-weight raising operations to $\mathbf{R}^{(W-1)}$. (See, e.g. Eqs. (4) and (5)).) Hence, $\mathbf{R}^{(W)}$, the space of special polynomials with scaling weight $W$, may be decomposed as follows:

$$\mathbf{R}^{(W)} = \left(q^{(1,1)}\mathbf{R}^{(W-1)}\right) \oplus \left(\partial_x \mathbf{R}^{(W-1)}\right) \oplus \tilde{\mathbf{R}}^{(W)} \ . \tag{8}$$

$\tilde{\mathbf{R}}^{(W)}$ contains special polynomials, which are not obtained by applying scaling-weight raising operators to $\mathbf{R}^{(W-1)}$.

Finally, $R^{(W,l)}$ may not be independent under linear combinations that involve operators, which raise or lower the scaling weight. For example,

$$R^{(3,2)} = 3\{R^{(3,1)} - \partial_x^{-1}(q^{(1,1)} R^{(3,1)})\} \quad , \quad R^{(4,4)} = 2\{R^{(4,2)} - \partial_x^{-1}(q^{(1,1)} R^{(4,1)})\} \ . \tag{9}$$

**2. From a single soliton to many solitons – a new perspective[1]**
It is customary to attribute the existence of soliton solutions of evolution equations to competition between nonlinear and dispersive terms. In this section, a new perspective is proposed, exploiting local special polynomials.

Consider the KdV equation (scaling weight, $W = 5$). When $u$ is a single-soliton solution, using the fact that $R^{(6,1)}[u] = 0$, one can reduce Eq. (1) to a first-order equation:

$$u_t = \{(2u^3 + (u_x)^3)/u^2\}u_x \ , \ (v = 4k^2) \ . \tag{10}$$

A similar reduction is obtainable for all equations with $W > 5$, which have the same single-KdV-soliton solution. In $W = 7$ there are two integrable equations: A symmetry of the KdV equation [1-22] and the Sawada-Kotera (SK) equation [24], given by

$$u_t = S_3^{KDV}[u] = u_5 + 10 u u_3 + 20 u_1 u_2 + 30 u^2 u_1 \ , \tag{11}$$

$$u_t = S_2^{SK}[u] = u_5 + 15 u u_3 + 15 u_1 u_2 + 45 u^2 u_1 \ . \tag{12}$$

Three equations have been identified in $W = 9$. Two are symmetries of the KdV and SK equations (hence, both integrable [1-22]), given, respectively, by

$$u_t = S_4^{KDV}[u] = u_7 + 14 u u_5 + 42 u_1 u_4 + 70 u_2 u_3 \\ + 70 u^2 u_3 + 280 u u_1 u_2 + 70(u_1)^3 + 140 u^3 u_1 \ , \tag{13}$$

---

[1] These ideas resulted from discussions with G.I. Burde, for which the author is indebted.

$$u_t = S_3^{SK}[u] = u_7 + 21uu_5 + 42u_1 u_4 + 63u_2 u_3$$
$$126u^2 u_3 + 378uu_1 u_2 + 63(u_1)^3 + 252u^3 u_1 \quad . \tag{14}$$

The third equation is the Caudrey-Dodd-Gibbon (CDG) equation, for which the existence of single-soliton and two-soliton solutions was demonstrated [26]:

$$u_t = S_2^{CDG}[u] = u_7 + 28uu_5 + 28u_1 u_4 + 70u_2 u_3$$
$$+ 210u^2 u_3 + 420uu_1 u_2 + 420u^3 u_1 \quad . \tag{15}$$

Eqs. (1), (11)-(15) have different multiple-solitons solutions, but share the same single-soliton solution. Exploiting the fact that local special polynomials vanish when $u$ is a single-soliton solution, Eqs. (1), (11)-(15) can be reduced for that case to

$$u_t = u_x \left\{ \left(2u^3 + (u_x)^2\right) / u^2 \right\}^n \quad , \quad v = (4k^2)^n \quad (W = 2n + 3) \quad . \tag{16}$$

Although it does not contain explicit dispersive terms, Eq. (16) is solved by the single-KdV-soliton solution of Eq. (2). However, multiple-solitons solutions of Eqs. (1), (11) - (15) obey Eq. (16) only asymptotically far from the soliton-collision region, where both right- and left-hand sides of Eq. (16) tend to

$$\sum_i v_i \partial_x u_{Single}[k_i] \quad \left(v_i = (4k_i^2)^n\right) \quad . \tag{17}$$

To obtain equations that have the same single-soliton solution, and, perhaps, also multiple-soliton solutions, one needs to add to Eq. (16) a term, which is written as:

$$u_t = u_x \left(\left(2u^3 + (u_x)^2\right) / u^2\right)^n + \left(R^{(6n+3)}[u]\right) / u^{2n} \quad . \tag{18}$$

$R^{(6n+3)}[u]$ is a polynomial of scaling weight $6n + 3$. In view of Eqs. (16) and (17), it must *vanish for a single-soliton solution, and be localized in the soliton-collision region when $u$ is a multiple-soliton solution*; hence, it has to be a local special polynomial.

For $W = 5$ ($n = 1$), $R^{(6n+3)}[u]$ has scaling weight 9. There are four linearly independent local special polynomials with $W = 9$ [36]:

$$R^{(9,1)} = uu_5 - u_2 u_3 + 21(u_1)^3 - 30u^3 u_1 \qquad R^{(9,2)} = u_1 u_4 - u_2 u_3 + 6(u_1)^3$$
$$R^{(9,3)} = u^2 u_3 - (u_1)^3 + 4u^3 u_1 \qquad R^{(9,4)} = uu_1 u_2 - (u_1)^3 + u^3 u_1 \tag{19}$$

Hence, for $n = 1$, Eq. (18) may be re-written as

$$u_t = \frac{\left(2u^3 + (u_x)^2\right)}{u^2} u_x + \frac{\left(a_1 R^{(9,1)}[u] + a_2 R^{(9,2)}[u] + a_3 R^{(9,3)}[u] + a_4 R^{(9,4)}[u]\right)}{u^2} \quad . \tag{20}$$

One now requires that Eq. (20) be a differential polynomial of scaling weight $W = 5$. This determines $a_1 - a_4$ uniquely and the KdV equation is resurrected.

In $W > 5$, this procedure leaves some freedom. Consider $W = 7$ ($n = 2$). The local special polynomial, $R^{(6n+3)}[u]$ in Eq. (19) now has scaling weight 15. There are 35 linearly independent local special polynomials with this weight! Requiring that Eq. (18) be a differential polynomial of $W = 7$, the coefficients of these polynomials, but one, are determined, leading to a more meaningful formulation of Eq. (18), which contains one free parameter:

$$u_t = S_3^{KDV}[u] + \mu R^{(7,1)}[u] \ . \tag{21}$$

$S_3^{KDV}$ is the $W = 7$ symmetry of the KdV equation (see Eq. (11)), and $R^{(7,1)}$ is the only local special polynomial in $W = 7$ (there are 36 non-local ones!), given by:

$$R^{(7,1)} = \partial_x R^{(6,1)} \ . \tag{22}$$

Eq. (21) is a one-parameter family of $W = 7$ equations, all of which have the same single-KdV-soliton solution. Up to trivial re-scaling, it is the most general family with this property. In higher scaling weights, such families will be multiple-parameter ones, because the number of linearly independent local special polynomials grows with $W$. In the $W = 5$ case, there was no ambiguity as there is no local special polynomial in $W = 5$.

The same ideas apply to other integrable evolution equations. For example, for a single-NLS-soliton solution, the NLS equation can be reduced to the following equation, which is solved by the single-soliton solution, but not by multiple-solitons solutions:

$$\psi_t = i\left((\psi^3 \psi^* + \psi_x^2)/\psi\right) \ . \tag{23}$$

With scaling weight assignments to $\psi$, $\partial_x$ and $\partial_t$ of, respectively, 1, 1 and 2, the NLS equation has $W = 3$. As there isn't a local special polynomial in $W = 3$, the NLS equation, like the KdV equation, is resurrected uniquely:

$$\psi_t = i\left(\frac{\psi^3 \psi^* + \psi_x^2}{\psi}\right) + i\left(\frac{R^{(4,1)}}{\psi}\right) = i\psi_{xx} + 2i\psi^2 \psi^* \ , \ \left(R^{(4,1)} = \psi^3 \psi^* + \psi \psi_{xx} - \psi_x^2\right). \tag{24}$$

$R^{(4,1)}$ is a local special polynomial of $W = 4$; it vanishes when is $\psi$ a single-NLS-soliton solution. When is $\psi$ a multiple-NLS-soliton solution, $R^{(4,1)}$ is localized in the soliton collision region. In higher scaling weights, multiple-parameters families of equations, which share the same single-NLS-soliton solution will emerge.

## 3. Special polynomials in the search for integrable equations [37]

Given an integrable evolution equation, with the aid of special polynomials, it is possible to find all the equations in higher scaling weights, which have the same single-soliton solution as the original equation and have, at least a two-solitons solution.

### 3.1 The KdV sequence

The KdV equation has $W = 5$. The first step is the construction of families of equations with $W > 5$, all of which have the same single-KdV-soliton solution. Up to trivial re-scaling, such equations are linear combinations of a symmetry [1-22] of the KdV equation (for which Eq. (2) is a solution) and all independent local special polynomials in the scaling weight considered. The $W = 7$ family is the single-parameter family given by Eq. (21). The $W = 9$ family has four free parameters:

$$u_t = S_4^{KDV}[u] + \alpha_1 R^{(9,1)}[u] + \alpha_2 R^{(9,2)}[u] + \alpha_3 R^{(9,3)}[u] + \alpha_4 R^{(9,4)}[u] \ . \tag{25}$$

$R^{(9,i)}$ are the four $W = 9$ linearly independent local special polynomials, given in Eq. (19).

The next step is the search for the equations within a family, which also have a two-solitons solution, similar in structure to the KdV-two-solitons solution, written as [23]:

$$\begin{aligned} u[t,x] &= 2\partial_x^2 \log f(t,x) \\ f(t,x) &= 1 + g(t,x,1) + g(t,x,2) + B\, g(t,x,1) g(t,x,2) \\ &\quad (g(t,x,i) = \exp\{2k_i(x + v_i t)\}) \end{aligned} \tag{26}$$

$B$ is not assigned its KdV value, to allow for the emergence of new equations.

In $W = 7$, Eq. (21) is solved also by a two-solitons solution only for $\mu = 0$ and $5$, corresponding, respectively, to the symmetry of the KdV equation, Eq. (11), and to the SK equation, Eq. (12) [24], both integrable.

In $W = 9$, Eq. (25) has also a two-solitons solution only for three sets of values of the coefficients, corresponding to the following equations. The first set is

$$\alpha_1 = \alpha_2 = \alpha_3 = \alpha_4 = 0 \ , \tag{27}$$

yielding the $W = 9$ symmetry of the KdV equation, Eq. (13). The second set is

$$\alpha_1 = 7, \alpha_2 = 0, \alpha_3 = 56, \alpha_4 = 98 \ , \tag{28}$$

yielding the $W = 9$ symmetry of the Sawada-Kotera equation, Eq. (14). The third set is:

$$\alpha_1 = 14 \ , \ \alpha_2 = -14 \ , \ \alpha_3 = \alpha_4 = 140 \ , \tag{29}$$

yielding the CDG equation, Eq. (15).

Eqs. (13) and (14) are integrable [26]. Eq. (15) was shown to have at least a two-solitons solution [26]. Thus, requiring that equations within the family of Eq. (25) have a two-solitons solution of the structure of Eq. (26), the already known integrable or potentially integrable equations emerge, and none else.

### 3.2 mKdV sequence

The mKdV equation ($W = 4$),

$$u_t = S_2^{MKDV} = 6u^2 u_1 + u_3 \ , \tag{30}$$

is integrable. Its soliton solutions are obtained through [27]

$$u(t,x) = 2\partial_x \tan^{-1}(g(t,x)/f(t,x)) \ . \tag{31}$$

The single-soliton solution is generated by

$$f(t,x) = 1 \qquad g(t,x) = q e^{k(x + k^2 t)} \ . \tag{32}$$

The two-solitons solution is generated by

$$f(t,x) = 1 - q_1 q_2 \left(\frac{k_1 - k_2}{k_1 + k_2}\right)^2 e^{k_1(x + k_1^2 t) + k_2(x + k_2^2 t)} \ , \quad g(t,x) = q_1 e^{k_1(x + k_1^2 t)} + q_2 e^{k_2(x + k_2^2 t)} \ . \tag{33}$$

In $W = 6$, the equations that have the same single soliton solution as Eq. (30) (with soliton velocity now equal to $k^4$) constitute a five-parameters family, given by

$$\begin{gathered} u_t = S_3^{MKDV}[u] + \alpha_1 R^{(6,1)} + \alpha_2 R^{(6,2)} + \alpha_3 R^{(6,3)} + \alpha_4 R^{(6,4)} + \alpha_5 R^{(6,5)} \\ \left(S_3^{MKDV}[u] = 30 u^4 u_1 + 10 u^2 u_3 + 40 u u_1 u_2 + 10(u_1)^3 + u_5\right) \end{gathered} \ . \tag{34}$$

In Eq. (34), $S_3^{MKDV}[u]$ is the $W = 6$ symmetry of the mKdV equation. The independent local special polynomials (vanishing when $u$ is a single-mKdV-soliton solution) are:

$$\begin{gathered} R^{(6,1)}[u] = \partial_x \left(u R^{(4,1)}[u]\right) \ , \quad R^{(6,2)}[u] = \partial_x^2 \left(R^{(4,1)}[u]\right) \ , \quad R^{(6,3)}[u] = u_1 R^{(4,1)}[u] \\ R^{(6,4)}[u] = u^2 R^{(4,1)}[u] \ , \quad R^{(6,5)}[u] = 7 u^3 u_2 + 2(u_2)^2 - 3 u_1 u_3 + u u_4 \\ \left(R^{(4,1)}[u] = u^4 - (u_1)^2 + u u_2\right) \end{gathered} \ . \tag{35}$$

To allow for the emergence of more than one equation that has two-solitons solutions, free coefficients in Eq. (33) were allowed:

$$\begin{gathered} f(t,x) = 1 + \sum_{k+l=2} \eta_{kl} \tilde{g}(t,x;1)^k \tilde{g}(t,x;2)^l \ , \quad g(t,x) = q_1 \tilde{g}(t,x,1) + q_2 \tilde{g}(t,x,2) \\ \left(\tilde{g}(t,x,i) = e^{k_i(x + v_i t)} \ , \quad v_i = k_i^4\right) \end{gathered} \ . \tag{36}$$

However, the only set of values of $\alpha_i$, $1 \leq i \leq 5$, for which Eq. (34) also has a two-solitons solution, is $\alpha_i = 0$; Eq. (34) is reduced to the symmetry of the mKdV equation:

$$u_t = S_3^{MKDV}[u] \ . \tag{37}$$

In $W = 8$, the equations that have a single-mKdV soliton solution constitute a 13-parameters family. However, again, the only equation that has both a single- and mKdV-like two-solitons solutions is the symmetry of the mKdV equation:

$$\begin{aligned} u_t = S_4^{MKDV}[u] &= 140 u^6 u_1 + 70 u^4 u_3 + 560 u^3 u_1 u_2 + 420 u^2 (u_1)^3 + 14 u^2 u_5 \\ &+ 84 u u_1 u_4 + 140 u u_2 u_3 + 126(u_1)^2 u_3 + 182 u_1 (u_2)^2 + u_7 \end{aligned} \tag{38}$$

### 3.3 The Kaup-Kupershmidt sequence
The Kaup-Kupershmidt (KK) equation ($W = 7$) [28-31],

$$u_t = S_2^{KK}[u] = u_5 + 30 u u_3 + 75 u_1 u_2 + 180 u^2 u_1 \ , \tag{39}$$

is integrable. Its solutions are obtained through the Hirota algorithm [23] as:

$$u[t,x] = \frac{1}{2} \partial_x^2 \log f(t,x) \ . \tag{40}$$

In the single-soliton case, $f(t,x)$ is given by

$$f(t,x) = 1 + g(t,x,1) + \frac{1}{16} g(t,x,1)^2 \ , \tag{41}$$

$$g(t,x,i) = \exp\{2 k_i (x + 16 k_i^4 t) + \delta_i\} \ . \tag{42}$$

The $W = 9$ family of equations that have the same single-KK-soliton solution is a two-parameter family, given by

$$u_t = S_3^{KK}[u] + \alpha_1 R^{(9,1)}[u] + \alpha_2 R^{(9,2)}[u] \ . \tag{43}$$

$S_3^{KK}[u]$ is the $W = 9$ symmetry of the KK equation,

$$\begin{aligned} S_3^{KK}[u] &= u_7 + 42 u u_5 + 147 u_1 u_4 + 252 u_2 u_3 + 504 u^2 u_3 \\ &+ 2268 u u_1 u_2 + 630 (u_1)^3 + 2016 u^3 u_1 \end{aligned} \tag{44}$$

The two linearly independent local special polynomials in $W = 9$ are

$$\begin{aligned} R^{(9,1)} &= -u_1 u_4 + u_2 u_3 + 6 u u_1 u_2 - 27(u_1)^3 + 48 u^3 u_1 \\ R^{(9,2)} &= 2 u u_5 + 3 u_1 u_4 - 5 u_2 u_3 + 60 u^2 u_3 + 60 u u_1 u_2 + 90(u_1)^3 \end{aligned} \tag{45}$$

To find the equations within this family, which also have a two-solitons solution, we search for solutions of the form

$$f(t,x) = 1 + g(t,x,1) + g(t,x,2)$$
$$+ A_1 g(t,x,1)^2 + A_2 g(t,x,2)^2 + B_1 g(t,x,1) g(t,x,2) \qquad (46)$$
$$+ B_2 g(t,x,1)^2 g(t,x,2) + B_3 g(t,x,1) g(t,x,2)^2 + B_4 g(t,x,1)^2 g(t,x,2)^2$$

Prospective equations correspond to specific values of $\alpha_1$ and $\alpha_2$ in Eq. (4) and of the coefficients in Eq. (46). The analysis yields that the only $W = 9$ equation, which has KK-type single- as well as two-solitons solutions, is the symmetry $S_3^{KK}[u]$, obtained for

$$\alpha_1 = \alpha_2 = 0 \quad . \qquad (47)$$

### 3.4 The bidirectional KdV (bKdV) sequence – an example of success
The bKdV equation is integrable [25, 26]. It is given by:

$$u_{tt} - u_{xx} - \partial_x S_2^{KdV}[u] = 0 \quad , \qquad (48)$$

with $S_2^{KdV}[u]$ defined in Eq. (1).

Eq. (48) has left- and right-moving soliton solutions. The solitons have the profiles of unidirectional KdV solitons, except for soliton velocity, which is given by

$$v_i = \pm\sqrt{1 + 4 k_i^2} \quad . \qquad (49)$$

$S_2^{KdV}[u]$ has $W = 5$. Hence, in Eq. (48), $\partial_x S_2^{KdV}[u]$ has $W = 6$. As there is one $W = 6$ local special polynomial, $R^{(6,1)}$ of Eq. (6), a question arises regarding the uniqueness of Eq. (48). Namely, are there other equations in this scaling weight, which have the potential to be integrable? To this end, consider the single-parameter family

$$u_{tt} - u_{xx} - \partial_x S_2^{KdV}[u] - \mu R^{(6,1)} = 0 \quad . \qquad (50)$$

Requiring that it have both single- and two-solitons solutions yields that $\mu = 0$ is the only possibility, so that Eq. (48) is the only integrable one in this scaling weight.

The sequence of higher scaling weight generalizations of Eq. (48) was identified in [26]. In the following, it is constructed through the exploitation of special polynomials.

We study the generalization to $W = 8$ and 10, for which the velocity of each bKdV soliton is given by

$$W = 8: \quad v_i = \pm\sqrt{1 + 16 k_i^4} \quad , \quad W = 10: \quad v_i = \pm\sqrt{1 + 64 k_i^6} \quad . \qquad (51)$$

Consider, first $W = 8$. The naïve extension of Eq. (48) is obtained by replacing $S_2^{KdV}[u]$ by its next symmetry, $S_3^{KdV}[u]$, given by Eq. (11):

$$u_{tt} - u_{xx} - \partial_x S_3^{KdV}[u] = 0 \ . \tag{52}$$

However, one readily finds that Eq. (52) has only a single-soliton solution, and does not have even a two-solitons solution. The question that arises naturally is whether special polynomials may help us to find equations that have, at least a two-solitons solution.

There are three independent local special polynomials in $W = 8$:

$$\begin{aligned} R^{(8,1)}[u] &= u\,R^{(6,1)}[u] \qquad & R^{(8,2)}[u] &= \partial_x^2 R^{(6,1)}[u] \\ R^{(8,3)}[u] &= u^2 u_2 - 3u(u_1)^2 + (u_2)^2 - u_1 u_3 & & \end{aligned} \tag{53}$$

Hence, the $W = 8$ equations, which have the same single-soliton solution as Eq. (52) constitute a three parameters family, given by

$$u_{tt} - u_{xx} - \partial_x S_3^{KdV}[u] - \alpha_1 R^{(8,1)}[u] - \alpha_2 R^{(8,2)}[u] - \alpha_3 R^{(8,3)}[u] = 0 \ . \tag{54}$$

Only one set of allowed values for $\alpha_i$ exists, for which Eq. (54) also has the two-solitons solution, of Eq. (26): $\alpha_1 = \alpha_3 = 0$ and $\alpha_2 = 5$. With these values, Eq. (54) becomes the bidirectional Sawada-Kotera (bSK) equation [26]:

$$u_{tt} - u_{xx} - \partial_x S_2^{SK}[u] = 0 \ , \tag{55}$$

where $S_2^{SK}[u]$ is given by Eq. (12).

The analysis in the case of $W = 10$ follows similar steps. Again, the naïve extension of Eq. (48),

$$u_{tt} - u_{xx} - \partial_x S_4^{KdV}[u] = 0 \ , \tag{56}$$

with $S_4^{KdV}[u]$ given in Eq. (13), has only a single-soliton solution. Again, special polynomials come to the rescue. There are seven linearly independent local special polynomials of $W = 10$. Hence, the equations that have the same single-soliton solution as Eq. (46) have the form:

$$u_{tt} - u_{xx} - \partial_x S_4^{KdV}[u] - \sum_{i=1}^{7} \alpha_i R^{(10,i)}[u] = 0 \ . \tag{57}$$

Requiring that Eq. (57) has a two-solitons solution, of the same structure as the solution of Eq. (48), one finds that this is possible only for one set of values of $\alpha_i$, $1 \le i \le 7$, for which it becomes the bCDG equation, [26]:

$$u_{tt} - u_{xx} - \partial_x S_2^{CDG}[u] = 0 \ . \tag{58}$$

$S_2^{CDG}[u]$ is given by Eq. (15).

Thus, the requirement of the existence of a two-solitons solution yields that the equations of the CDG hierarchy [26] are the only possible higher-scaling weight extensions of Eq. (48). However, this procedure does not amount to a proof of integrability. In fact, Eqs. (55) and (58) are known to have at least a two-solitons solution, however, it is not known whether they are integrable [26].

**3.5 The GKK soliton – an example of failure**
In a mixed scaling weight equation (KdV + a $W = 7$ term), a new type of a single-soliton solution, the generalized (GKK) soliton, has been found [39-42]. That equation was shown to have also two- and three-GKK solitons solutions [39]. In pure scaling weights, the GKK-soliton solution appears for the first time in $W = 9$. However, its equation does not have any multiple-GKK soliton solutions.

Consider the most general $W = 9$ equation in the KdV sequence:

$$u_t = S[u] = a_1 u^3 u_1 + a_2 u^2 u_3 + a_3 u u_1 u_2 + a_4 (u_1)^3 \\ + a_5 u u_5 + a_6 u_1 u_4 + a_7 u_2 u_3 + u_7 \quad . \tag{59}$$

We look for a single-soliton solution, using the Hirota transformation [23] augmented by the expansion procedure developed in [43]:

$$u = M \partial_x^2 \log[f(t,x)] \\ f(t,x) = 1 + \varepsilon f_1(\xi) + \varepsilon^2 f_2(\xi) \quad . \tag{60} \\ (\xi = x + 64 k^2 t)$$

Requiring that Eq. (55) be obeyed order-by order in $\varepsilon$, one obtains through $O(\varepsilon^2)$:

$$f_1(\xi) = q e^{2k\xi} \\ f_2(\xi) = \left\{ \frac{1}{2} - \frac{(a_5 + a_6 + a_7)}{504} M \right\} q^2 e^{4k\xi} \quad . \tag{61}$$

Proceeding to higher orders in $\varepsilon$, and setting $\varepsilon = 1$ at the end of the analysis, one finds that the coefficients in Eq. (59) correspond to four possibilities. Three of them correspond to the single-KdV, SK or KK solitons. The fourth possibility is the single-GKK soliton. Two of the coefficients are free, chosen here to be $a_5$ and $a_6$, and the others are given by:

$$a_1 = \frac{8820}{M^3} - \frac{206 a_5}{M^2} - \frac{76 a_6}{M^2} + \frac{14 a_5^2}{9M} + \frac{8 a_5 a_6}{9M} + \frac{8 a_6^2}{63 M}$$
$$a_2 = \frac{14 a_5^2}{27} + \frac{8 a_5 a_6}{27} + \frac{8 a_6^2}{189} + \frac{1743}{M^2} - \frac{295 a_5}{6M} - \frac{55 a6}{3M}$$
$$a_3 = -\frac{35 a_5^2}{27} - \frac{20 a_5 a_6}{27} - \frac{20 a_6^2}{189} - \frac{2562}{M^2} + \frac{461 a_5}{3M} + \frac{103 a6}{3M} \tag{62}$$
$$a_4 = \frac{7 a_5^2}{9} + \frac{4 a_5 a_6}{9} + \frac{4 a_6^2}{63} + \frac{2016}{M^2} - \frac{94 a_5}{M} - \frac{14 a6}{M} \quad , \quad a_7 = -\frac{11 a_5}{4} - \frac{3 a_6}{2} + \frac{693}{2M}$$

Substituting the expression for $a_7$ in Eqs. (60) and (61) the new single-soliton solution is found:

$$u(t,x) = \frac{4k^2 M Q\{Q + \cosh[2k(x + 64k^2 t) + \delta]\}}{\{1 + Q\cosh[2k(x + 64k^2 t) + \delta]\}^2} \quad . \tag{63}$$

$$Q = \frac{1}{6}\sqrt{\frac{(7a_5 + 2a_6)M - 378}{14}}, \qquad e^\delta = q\frac{Q}{2}$$

This solution is reduced single-KdV soliton solution when $Q \to 0$, and to the single-KK soliton solution − when $Q = 1$. Examples of single GKK-solitons are shown in Fig. 3.

To find whether Eq. (59) has also a two GKK-solitons solution, one needs to perform the expansion of [43] through $O(\varepsilon^4)$:

$$f(t,x) = 1 + \varepsilon f_1(\xi_1,\xi_2) + \varepsilon^2 f_2(\xi_1,\xi_2) + \varepsilon^3 f_3(\xi_1,\xi_2) + \varepsilon^4 f_4(\xi_1,\xi_2)$$
$$(\xi_i = x + 64k_i^2 t, \qquad i = 1,2) \tag{64}$$

It is found that there is no a version of Eq. (59), which has the GKK single-soliton solution, and, at least, a two-solitons solution. The coefficients $a_5$ and $a_6$ cease to be free, and obtain values, which reduce Eq. (59) into symmetries of the KdV, SK and KK equations, all integrable, or to the CDG equation, the integrability of which has not been established yet.

**4. Special polynomials in perturbed evolution equations**
Integrable equations provide approximations to complex dynamical systems. The remainder of the equations of the original system generates a perturbation, characterized by a small parameter, |ε| « 1. The role played by special polynomials in the analysis of solutions has been studied in the cases of the perturbed KdV, mKdV and NLS equations, using the Normal-Form expansion method [13, 16, 32-34]. The same picture emerges in all cases. Here, only the case of the perturbed KdV equation is reviewed [35, 36, 38].

**4.1 Special polynomials in the analysis of the perturbed KdV equation**
The generic form of the perturbed KdV equation is:

$$\begin{aligned}w_t = {}& 6ww_1 + w_3 \\ & + \varepsilon\left(30\alpha_1 w^2 w_1 + 10\alpha_2 ww_3 + 20\alpha_3 w_1 w_2 + \alpha_4 w_5\right) \\ & + \varepsilon^2 \begin{pmatrix} 140\beta_1 w^3 w_1 + 70\beta_2 w^2 w_3 + 280\beta_3 w w_1 w_2 + 14\beta_4 w w_5 \\ + 70\beta_5 w_x^3 + 42\beta_6 w_1 w_4 + 70\beta_7 w_2 w_3 + \beta_8 w_7 \end{pmatrix} + O(\varepsilon^3)\end{aligned} \tag{57}$$

(A detailed analysis has been performed through $O(\varepsilon^3)$ in {36], [38].)

Expanding $w$ in an asymptotic series:

$$w(t,x) = u(t,x) + \varepsilon u^{(1)}(t,x) + \varepsilon^2 u^{(2)}(t,x) + O(\varepsilon^3) \quad . \tag{58}$$

The zero-order term, $u$, is governed by the Normal Form [13, 16, 19, 20, 32-34], which has the same $N$-solitons solutions, for any $N$ as the unperturbed KdV equation:

$$u_t = S_2^{KdV}[u] + \varepsilon\, \alpha_4\, S_3^{KdV}[u][u] + \varepsilon^2\, \beta_8\, S_4^{KdV}[u] + O(\varepsilon^3) \ . \tag{59}$$

Hence, $u(t,x)$ represents elastic soliton scattering. The only effect of Eq. (59) is to update the soliton velocities according to [1-20]

$$v_i = 4\, k_i^2 + \varepsilon\alpha_4\, 16\, k_i^4 + \varepsilon^2\, \beta_8\, 64\, k_i^4 + O(\varepsilon^3) \ . \tag{60}$$

### 4.2 Elastic and inelastic components

When $u$ is a multiple-solitons solution ($N > 1$), each correction term, $u^{(n)}$, in Eq. (58) can be decomposed into a sum of an elastic component and an inelastic one. The elastic component has the functional form obtained for $u^{(n)}$ in the single-soliton case. It preserves the elastic nature of soliton collisions, characterizing the zero-order term, $u$.

The inelastic component represents purely inelastic effects of soliton interactions, induced by the perturbation. *Both the part of the perturbation, which generates inelastic interactions, as well as the inelastic component in the solution, can be cast in terms of special polynomials.* This will be shown in detail for the first-order correction, $u^{(1)}$, which has been constructed in [32]. The result of [32] is re-written as:

$$u^{(1)} = u^{(1)}_{el} + u^{(1)}_{in} \ . \tag{61}$$

The elastic component is given by [36, 38]

$$u^{(1)}_{el} = \left(-\tfrac{5}{2}\alpha_1 + \tfrac{10}{3}\alpha_2 + \tfrac{5}{3}\alpha_3 - \tfrac{5}{2}\alpha_4\right) u_2 + \left(-5\alpha_1 + 5\alpha_2\right)(u)^2 + S \ . \tag{62}$$

$S$ is a linear combination of symmetries of the KdV equation, facilitating satisfaction of initial or boundary data. The elastic component contains only powers of $u$ and of $\partial_x^k u$. As a result, it retains the elastic nature of solitons collisions that characterizes $u$.

The inelastic component, $u^{(1)}_{in}$, is given by [36, 38]

$$u^{(1)}_{in} = -\tfrac{10}{3}\left(\alpha_2 - \alpha_4\right) R^{(4,1)}[u] \ . \tag{63}$$

As $u^{(1)}_{in}$ is proportional to the non-local special polynomial, $R^{(4,1)}$ (see Eq. (4)), it is of a purely inelastic nature. The amplitude of its contribution along each soliton trajectory depends on the wave numbers of *all other* solitons. (See, e.g., Eq. (7)).

$u^{(1)}_{in}$ of Eq. (63) is the solution of the first-order equation:

$$\partial_t u^{(1)}_{in} = 6\partial_x\left(u\, u^{(1)}_{in}\right) + \partial_x^3 u^{(1)}_{in} + 10\left(\alpha_2 - \alpha_4\right) R^{(7,1)}[u] \ . \tag{64}$$

The driving term is the local special polynomial, $R^{(7,1)}$ (see Eqs. (22) and (6)).

## 4.3 "Obstacles to asymptotic integrability" and local special polynomials

In higher orders, the part of the perturbation, which drives $u_{in}^{(n)}$, $n \geq 2$, can be also cast in the form of a local special polynomial. For example, $u_{in}^{(2)}$, the inelastic contribution in $u^{(2)}$, is driven by a linear combination of the $W = 9$ local special polynomials of Eq. (19).

The solution of Eq. (64) for $u_{in}^{(1)}$ is the closed-form expression of Eq. (63) - a non-local special polynomial. In orders $n \geq 2$, most inelastic driving terms also generate closed-from contributions to $u_{in}^{(n)}$ in the form of non-local special polynomials. However, some driving terms generate contributions, for which differential polynomial expressions cannot exist. These terms are "obstacles to asymptotic integrability" [13, 16, 33, 34].

Consider, for example, the case of the second-order correction to the solution. The equation for $u_{in}^{(2)}$ can be written as:

$$\partial_t u_{in}^{(2)} - 6 \partial_x \left( u u_{in}^{(2)} \right) - \partial_x^3 u_{in}^{(2)} = $$
$$\partial_x \left\{ \begin{array}{l} \mu_1 R^{(8,1)}[u] \\ + \mu_2 \left( 3 R^{(8,1)}[u] + R^{(8,2)}[u] + R^{(8,3)}[u] \right) + \mu_3 \left( 2 R^{(8,1)}[u] + R^{(8,3)}[u] \right) \end{array} \right\} . \quad (65)$$

The driving term is a $W = 9$ local special polynomial, expressed in terms of the $W = 8$ local special polynomials of Eq. (53). The coefficients, $\mu_j$, are known combinations of the unspecified coefficients in Eq. (57).

The terms proportional to $\mu_2$ and $\mu_3$ generate closed-form, non-local special-polynomial contributions to $u_{in}^{(2)}$. For example, the contribution of the $\mu_3$-term to the solution of Eq. (65), shown in the following (together with its asymptotic form in the two-solitons case) is:

$$\mu_3 \partial_x \left\{ -\left( q^{(3,1)} u + \frac{7}{3} u u_1 + \frac{1}{3} u_3 \right) \right\} \xrightarrow[t \to \pm\infty]{} \pm \mu_3 \partial_x \left\{ \frac{8}{3} k_1^3 u_{Single}[k_2] - \frac{8}{3} k_2^3 u_{Single}[k_1] \right\} . \quad (66)$$

Eq. (66) demonstrates the inelastic nature of non-local special polynomials: For $|t| \to \infty$, the contribution of a soliton is affected by the wave number of the *other* soliton.

However, the term proportion to $\mu_1$ in Eq. (65) is an obstacle if $\mu_1 \neq 0$. This can be seen as follows. Denoting the contribution of the by $\mu_1$ – term to $u_{in}^{(2)}$ by $\mu_1 \partial_x \omega^{(2)}$, the equation for $\omega^{(2)}$ is:

$$L \omega^{(2)} = R^{(8,1)}[u]$$
$$\left( L \equiv \partial_t - 6 u \partial_x - \partial_x^3 \right) \quad (67)$$

$R^{(8,1)}$, the driving term in Eq. (67), is a $W = 8$ local special polynomial. The differential operator, $L$, has scaling weight 3. Hence $\omega^{(2)}$ must have $W = 5$. $\mathbf{R}^{(5)}$, the space of all $W = 5$ special polynomials, is a finite dimensional linear space (there are 10 linearly independent special polynomials in this scaling weight). A direct check yields that $R^{(8,1)}$ is not contained in $L \bullet \mathbf{R}^{(5)}$. Hence, $\omega^{(2)}$ cannot be written in closed form as a differential polynomial in $u$, and has to be found numerically. Fig. 2 shows the numerical solution for $\omega^{(2)}$, when $u$ is a two-solitons solution. Zero-initial data were chosen for large negative time.

$\omega^{(2)}$ contains contributions along the two solitons, similar to the closed-form ones generated by the other driving terms in Eq. (65). (See, e.g., Eq. (66).) Apart from the numerically determined amplitude, the profile of the asymptotic contribution along each soliton is identical to that of the same soliton.

However, $\omega^{(2)}$ also contains a dispersive wave. The localized nature of the special polynomial, $R^{(8,1)}$, and of the solitons, allows one to obtain the following approximate equation for $\omega^{(2)}$:

$$\partial_t \omega^{(2)} \cong \partial_x^3 \omega^{(2)} \quad (x \gg 0, t \gg 0, x + k_1 t \gg 0, x + k_2 t \gg 0) \ . \tag{68}$$

Eq. (68) is a valid approximation away from the soliton-collision region and from the two solitons. A solution of Eq. (68) using of Hypergeometric functions mimics the dispersive wave.

## 5. Open problems
The ideas presented in the previous sections raise two questions.

1. <u>Conditions for integrability</u>: In the cases studied here, requiring the existence of single- and two-solitons solutions was sufficient for identifying, in a given scaling weight, all the equations that are integrable, or, at least have single-soliton and two-solitons solutions. Are there general conditions, under which the existence of these two solutions is sufficient to ensure that a nonlinear evolution equation is integrable, and, in particular, has $N$-soliton solutions for any $N > 1$? Is the special role played by the two-solitons solution connected in any way with the fact that the Hirota construction of multiple-soliton solutions [23] is based solely on two-wave interactions?

2. <u>Symmetry properties of special polynomials</u>: Let $n$ the order in the expansion, in which an obstacle to asymptotic integrability appears for the first time. Similar to Eqs. (64) and (65), in all orders, $m \leq n$, the equation obeyed by the inelastic component, is :

$$L u_{in}^{(m)} \equiv \partial_t u_{in}^{(m)} - 6 \partial_x \left( u u_{in}^{(m)} \right) - \partial_x^3 u_{in}^{(m)} = R^{(2m+5)}[u] \quad , \quad m \leq n \ . \tag{69}$$

The driving term, $R^{(2m+5)}$, is a special polynomial of $W = 2m + 5$. The differential operator $L$ in Eq. (69) has $W = 3$. Hence, $u_{in}^{(m)}$ must have $W = 2m + 2$. The dimensions of the spaces of special polynomials grow with $W$,

$$\dim \left\{ \mathbf{R}^{(W=2m+2)} \right\} < \dim \left\{ \mathbf{R}^{(W=2m+5)} \right\} \ . \tag{70}$$

There is no polynomial solution for $u_{in}^{(n)}$ if $R^{(2n+5)}$ is not contained in the image of $\mathbf{R}^{(W=2n+2)}$ under $L$:

$$R^{(2n+5)}[u] \notin L \mathbf{R}^{(W=2n+2)} \ . \tag{71}$$

If Eq. (8), the decomposition of $\mathbf{R}^{(W)}$, were associated with the irreducible representations of some symmetry group, this would contribute to understanding when a perturbed integrable equation is or is not asymptotically integrable.

**Appendix I. Bounded non-local entities**

The first few $q^{(W,l)}$ are:

$\underline{W=1}$
$$q^{(1,1)} = \partial_x^{-1} u \tag{I.1}$$

$\underline{W=3}$
$$q^{(3,1)} = \partial_x^{-1}(u^2) \tag{I.2}$$

$\underline{W=4}$
$$q^{(4,1)} = \partial_x^{-1}(u^2 q^{(1,1)}) \tag{I.3}$$

$\underline{W=5}$
$$q^{(5,1)} = \partial_x^{-1}(u^3) \quad , \quad q^{(5,2)} = \partial_x^{-1}\left(u^2 \left(q^{(1,1)}\right)^2\right) \quad , \quad q^{(5,3)} = \partial_x^{-1}\left((u_1)^2\right) \tag{I.4}$$

$\underline{W=6}$
$$q^{(6,1)} = \partial_x^{-1}(u^3 q^{(1,1)}) \quad , \quad q^{(5,2)} = \partial_x^{-1}\left(u^2 \left(q^{(1,1)}\right)^3\right) \quad , \quad q^{(5,3)} = \partial_x^{-1}\left((u_1)^2 q^{(1,1)}\right) \tag{I.5}$$

To guarantees that non-local $R^{(W,l)}$ vanish when $u$ is a single-soliton solution, and represent pure inelastic contributions along soliton trajectories when $u$ is a multiple-solitons solution, the $\partial_x^{-1}$ operation has to be defined by

$$\partial_x^{-1} f(t,x) \equiv \frac{1}{2}\left(\int_{-\infty}^{x} f(t,x) dx - \int_{x}^{\infty} f(t,x) dx\right) . \tag{I.6}$$

**Appendix II. Computing special polynomials**

The procedure for the construction of special polynomials is straightforward. One constructs the most general differential polynomial in $u$, the solution of an evolution equation, in a give scaling weight, $W$. The polynomial may contain monomials comprised of powers of $u$ and $\partial_x^k u$, as well as bounded non-local entities, examples of which have been presented in Appendix I. Each polynomial is multiples by an unspecified coefficient. One then substitutes for $u$ the single-soliton solution, and requires that the differential polynomial vanish identically. This determines some of the coefficients. The number of undetermined coefficients equals the number of linearly independent special polynomials in the scaling weight considered.

AS an example, consider the case of $W = 3$, with $u$ being the solution of the KdV equation. The most general differential polynomial in this scaling weight is:

$$R^{(3)} = a_1 u_x + a_2 q u + a_3 q^3 + a_4 q^{(3,1)} . \tag{II.1}$$

Substituting the single-KdV soliton solution in $R^{(3)}$, and requiring that it vanish identically reveals that two of the coefficients remain undetermined, so that there are only two linearly independent special polynomials, the ones given by Eq. (3).

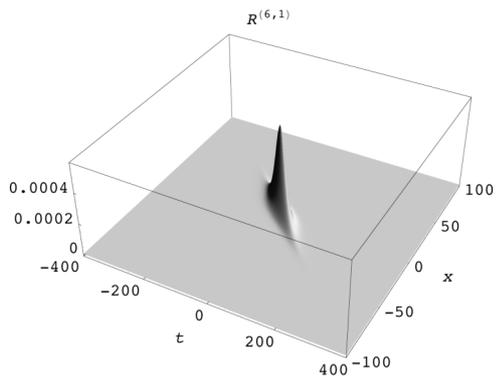

Fig. 1 Local special polynomial $R^{(6,1)}$; $u$ is a two-solitons solution; $k_1 = 0.25$, $k_2 = 0.15$.

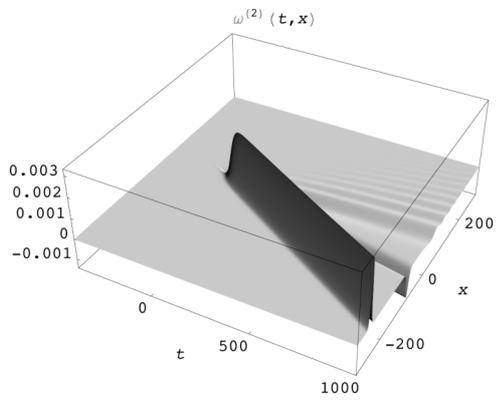

Fig.2 Solution of Eq. (67); $u$ is a two-solitons solution; $k_1 = 0.25$, $k_2 = 0.15$.

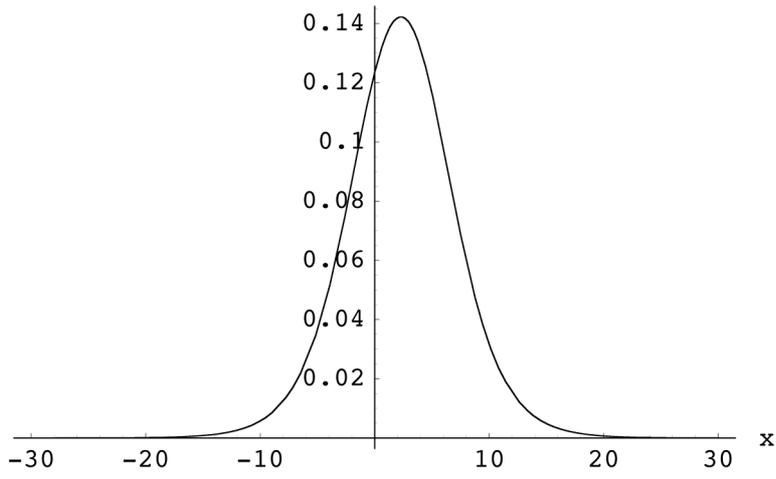

Fig. 3a GKK single-soliton with $k = 0.2$, $Q = 0.8$

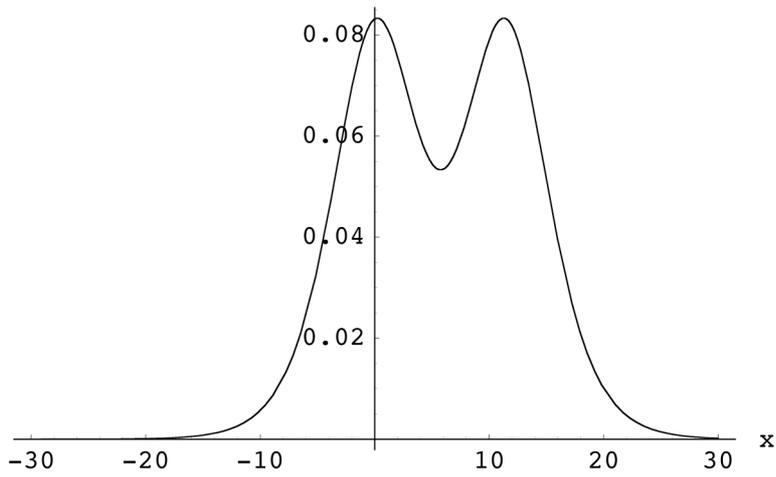

Fig. 3b GKK single-soliton with $k = 0.2$, $Q = 0.2$